\documentclass{article}
\usepackage[T1]{fontenc} % add special characters (e.g., umlaute)
\usepackage[utf8]{inputenc} % set utf-8 as default input encoding
\usepackage{ismir,amsmath,cite,url}
\usepackage{graphicx}
\usepackage{color}
\usepackage{amsmath,amssymb,multirow, booktabs, subcaption}

\usepackage{lineno}
\title{Emotion Embedding Spaces for Matching Music to Stories}

\multauthor
{
	Minz Won$^{1}$  \hspace{.35cm} 
	Justin Salamon$^{2}$  \hspace{.35cm} 
	Nicholas J. Bryan$^{2}$  \hspace{.35cm} 
	Gautham J. Mysore$^{2}$  \hspace{.35cm} 
	Xavier Serra$^{1}$  \hspace{.35cm} 
} 
{
	$^1$ Music Technology Group, Universitat Pompeu Fabra, Barcelona, Spain\\
	$^2$ Adobe Research, San Francisco, California, United States\\
	{\tt\small minz.won@upf.edu}
}
\def\authorname{Minz Won, Justin Salamon, Nicholas J. Bryan, Gautham J. Mysore, and Xavier Serra}

\begin{document}

\maketitle
\begin{abstract}
Content creators often use music to enhance their stories, as it can be a powerful tool to convey emotion. In this paper, our goal is to help creators find music to match the emotion of their story. We focus on text-based stories that can be auralized (e.g., books), use multiple sentences as input queries, and automatically retrieve matching music. We formalize this task as a cross-modal text-to-music retrieval problem. Both the music and text domains have existing datasets with emotion labels, but mismatched emotion vocabularies prevent us from using mood or emotion annotations directly for matching. To address this challenge, we propose and investigate several emotion embedding spaces, both manually defined (e.g., valence/arousal) and data-driven (e.g., Word2Vec and metric learning) to bridge this gap. 
Our experiments show that by leveraging these embedding spaces, we are able to successfully bridge the gap between modalities to facilitate cross modal retrieval. 
We show that our method can leverage the well established valence-arousal space, but that it can also achieve our goal via data-driven embedding spaces. By leveraging data-driven embeddings, our approach has the potential of being generalized to other retrieval tasks that require broader or completely different vocabularies.

% Experimental results show that the well known, manually defined, valence/arousal space suits the task. Furthermore, our proposed three-branch metric learning model can integrate the process into a fully data-driven pipeline, by leveraging heterogeneous mood taxonomies while preserving emotional continuity in its embedding space.

% Our experiments show that by leveraging these embedding spaces, we are able to facilitate cross modal retrieval. We show that the well established valence-arousal space can be used to bridge between modalities, but that can be also achieved with a fully data-driven embedding space that we propose. Our data-driven embedding space successfully leverages heterogeneous mood taxonomies to form a continuous emotion space that facilitates cross-modal text-to-music retrieval.

% Our experimental results show that the well known, manually defined, valence/arousal space is a powerful model, but that we are able to obtain similarly good retrieval results via learned, data-driven embeddings. Furthermore, the retrieval model can be trained in a single end-to-end scheme when using metric learning to learn the emotion embedding space so that it can preserve neighboring structure within modalities and enable emotional continuity in it.
% and enable a wider variety of cross-modal retrieval applications in the future.
\end{abstract}

\section{Introduction}\label{sec:introduction}
\begin{figure}
 \centerline{
 \includegraphics[width=1.00\columnwidth]{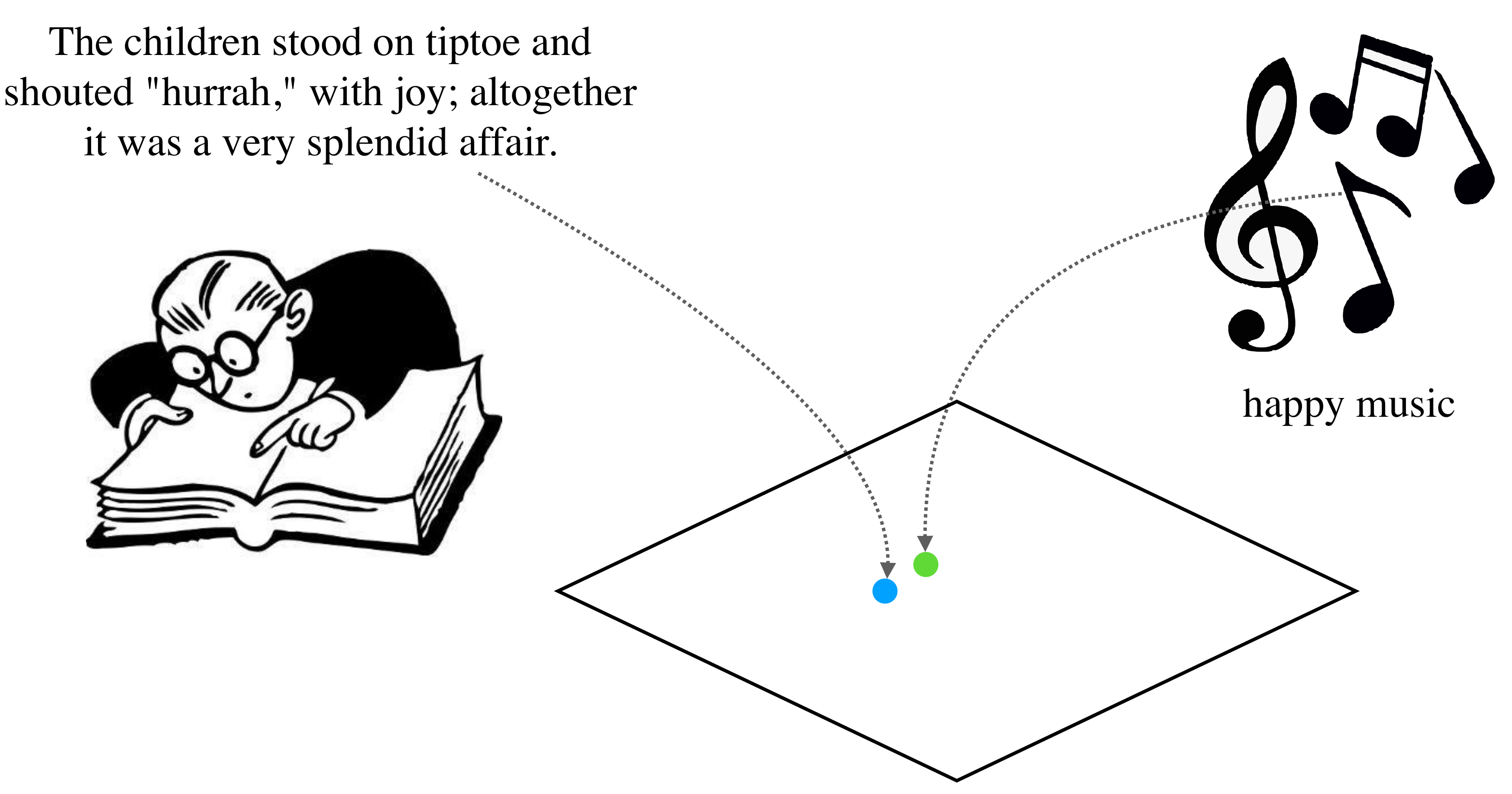}}
 \caption{Cross-modal text-to-music retrieval using an aligned, multimodal embedding space.}
 \label{fig:dist}
\end{figure}

% Motivation + problem
Content creators, both amateur and professional alike, often use music to enhance their storytelling due to its powerful ability to elicit emotion~\footnote{We use the terms \textit{emotion} and \textit{mood} interchangeably following previous work~\cite{kim2010music}.}. For example, when dissonant music is added to a horror movie, it can amplify the scary mood of the story line. Similarly, cheerful music can emphasize the excited mood in a scene of a birthday party. Matching text and music to create a narrative, typically requires tediously browsing large-scale music collections,  significant experience, and musical expertise. In this paper, we therefore address the problem of automatically matching music to text as shown in Figure~\ref{fig:dist}. 

We formalize this task as a cross-modal retrieval problem~\cite{wang2017adversarial} and focus on matching long-form text (multiple sentences, paragraphs) to music. For queried sentences like books and scripts, we seek to retrieve matching music for applications such as podcasts, audio books, movies, and film. To facilitate cross-modal retrieval, a common approach is to first perform feature extraction to convert each data modality into an embedding space. Then, the different embedding spaces must be matched to bridge the \textit{modality gap} by somehow aligning their different distributions~\cite{wang2017adversarial}. Once aligned, (fast) nearest neighbor search can be used for retrieval.

%Also, this can be expanded towards films where textual information plays an important role (i.e., scripts) along side with video.

% Music evokes emotions and the conveyed emotions are highly consistent within the same culture. Sometimes music is used in movies or audio books to inject more dramatic effects to the story line. To accomplish this, traditionally, trained experts understand the given contents and choose appropriate music based on their interpretation of the contents. However, this requires long training to develop expertise and itself is a time-consuming process that includes reading or watching the contents, and browsing music. In this paper, our main motivation is to automate this content-to-music retrieval process, so that non-experts can easily create their contents with appropriate music and also experts can browse suitable music in more efficient ways. Especially, our main focus is on the text-to-music retrieval problem using books and scripts.

% \subsection{Problem Definition}

% Different methods for cross-modal retrieval, but it's not that simple
Various methods have been proposed for cross-modal feature extraction and alignment. For example, canonical correlation analysis has been used to bridge the modality gap~\cite{yao2015learning} as well as modern deep learning techniques that learn common representation spaces~\cite{zhen2019deep,wei2016cross}. Such methods can be categorized into four groups: unsupervised, pairwise-based, rank-based, and supervised methods~\cite{wang2016comprehensive}. Among these, supervised methods are the most straightforward and in theory can take advantage of existing labeled datasets (e.g., labels of \textit{happy}, \textit{sad}) and themes (e.g., \textit{party}, \textit{wedding} with corresponding text and music). Difficulties, however, immediately arise because of mismatched dataset taxonomies (vocabularies) per modality, making it challenging to use standard techniques directly.

% Our solution
Therefore, in this work we focus on the task of emotion-based text- (e.g. sentences, paragraphs) to-music retrieval, and investigate how we can best perform cross-modal retrieval with heterogeneous dataset taxonomies. To the best of our knowledge, this problem has not been previously addressed and could be beneficial to media content creation applications. 
We propose six different deep learning strategies to extract relevant features and bridge the modality gap between text and music including (1) classification (2) multi-head classification (3) valence-arousal regression (4) Word2Vec regression (5) two-branch metric learning and (6) three-branch metric learning.
We then evaluate each approach on multiple text and music datasets, report objective results via precision at five and mean reciprocal rank, and conclude with qualitative analysis and discussion. Our results show that our valence-arousal-based method is a powerful baseline for emotion-based cross-modal retrieval, but that our three-branch metric-learning approach is comparable, more general, and does not require manually engineered valence and arousal mappings.

\section{Related Work}\label{sec:related}

%%%%%%%%%%%%%%%%%%%%%%%%%%%%%%%%%%%%%%%%%%%%%%%%%%%%%%%%%%%%%%%%%%%%%%%%%%%%%%%%
%%%%%%%%%%%%%%%%%%%%%%%%%%%%%%%%%%%%%%%%%%%%%%%%%%%%%%%%%%%%%%%%%%%%%%%%%%%%%%%%
\subsection{Text Emotion Classification}
Text emotion classification methods or the task of predicting emotion from text can be divided into three categories: lexicon-based models, traditional machine learning models, and deep learning models. 
% Text emotion recognition models can be divided into three categories: lexicon-based models~\cite{mohammad2013crowdsourcing,strapparava2004wordnet}, traditional machine learning models~\cite{danisman2008feeler,hasan2014emotex}, and modern deep learning models~\cite{abdul2017emonet,batbaatar2019semantic,cortiz2021exploring}. 
Lexicon-based models take advantage of pre-defined emotion lexicons, such as NRC EmoLex~\cite{mohammad2013crowdsourcing} and WordNet-Affect~\cite{strapparava2004wordnet} to match keywords. Traditional machine learning approaches recognize emotions using algorithms such as support vector machine (SVM)~\cite{danisman2008feeler} and Naive Bayes~\cite{hasan2014emotex}. Finally, deep learning models use deep sequence models such as gated recurrent unit (GRU)~\cite{abdul2017emonet}, bidirectional long-short term memory (BiLSTM)~\cite{batbaatar2019semantic}, and Transformers~\cite{cortiz2021exploring}. Most recently, Transformer models~\cite{devlin2018bert,liu2019roberta,sanh2019distilbert} have become quite prevalent. Such models take  advantage of transfer learning, are commonly pre-trained to learn language representation with large datasets, and then applied to various downstream tasks including question and answer systems as well as emotion recognition~\cite{cortiz2021exploring}.
\subsection{Music Emotion Classification}
Music emotion classification or the task of predicting emotion from music audio is commonly divided into conventional feature extraction and prediction approaches~\cite{tzanetakis2007marsyas,peeters2008generic,cao2009thinkit}, and end-to-end deep learning approaches~\cite{lidy2016parallel, delbouys2018music}. Deep learning approaches have become most prevalent and commonly frame emotion recognition as a multi-class or multi-label auto-tagging classification problem~\cite{choi2016automatic,lee2017sample, pons2018end, won2020data, lee2020metric}. 
%Music tagging is a bigger umbrella that subsumes music emotion recognition that aims at predicting various music tags such as genre and instrument as well as emotion. 
Recently, multiple music tagging models were evaluated in a homogeneous evaluation pipeline~\cite{won2020evaluation} and found three design recommendations for automatic music tagging models: (1) use mel-spectrogram inputs, (2) use $3\times3$ convolutional filters, and (3) use short-chunk audio inputs with small hop sizes and max-pooling. Based on this, a model using mel-spectrogram inputs and convolutional neural networks with focal loss~\cite{lin2017focal} won the MediaEval 2020 Emotion-and-Theme-Recognition-in-Music-Task\footnote{https://multimediaeval.github.io/2020-Emotion-and-Theme-Recognition-in-Music-Task}~\cite{knox2020media}. 
% In our work, we use this architecture with residual connections for music analysis.

% \subsection{Music Representation Learning}
% Music tagging is a bigger umbrella that subsumes music emotion recognition. The goal of music tagging is to predict relevant tags, not only emotion but also instrument or genre. Previous music tagging research have actively explored different music representation models using convolutional neural networks~\cite{choi2016automatic,pons2018end,lee2017sample,won2020data} and they are recently evaluated in a homogeneous evaluation pipeline~\cite{won2020evaluation}. Two main take home from the evaluation work are to 1) use Mel spectrogram inputs and 2) stay with $3\times3$ convolutional filters instead of arbitrary design.

% Again, we use a pretrained model.

% For music representation learning, many different architectures have been proposed. Based on previous work~\cite{won2020evaluation}, we use a simple but the best short-chunk ResNet. We pretrain it on MTAT. We used MTAT for pretraining since we considered MSD and MTG-Jamendo as our dataset as we specified in the previous section (data selection).

% For mood prediction, it is known that Focal loss is critical (winner of mood classification challenge MediaEval 2020) but we do not use it because Audioset is already balanced.

\subsection{Valence-arousal Regression \& Word Embeddings}
Beyond classification, previous works~\cite{schmidt2010feature,han2009smers} suggest that regression approaches can outperform classification approaches in music emotion recognition. Here, researchers use the well-known valence-arousal emotion space~\cite{russell1980circumplex,thayer1990biopsychology} where valence represents positive-to-negative emotions, and arousal indicates the intensity of the emotions. These annotations can be collected by human annotators directly~\cite{schmidt2010feature} or by mapping existing mood labels into the valence-arousal space using pre-defined lexicons~\cite{delbouys2018music,mohammad2018obtaining}. 
%With the latter mapping process, multiple datasets with different taxonomies can be bridged together in the same 2D space which facilitates nearest neighbor search of relevant items. (make a contribbution, not related work)

% \subsubsection{Word embedding}
% Pretrained Word2Vec
\begin{table}
\centering
\footnotesize
\begin{tabular}{|@{\hskip0.8pt}c@{\hskip0.8pt}|@{\hskip0.8pt}c@{\hskip0.8pt}|@{\hskip0.8pt}c@{\hskip0.8pt}|}
\hline
Tag  & GoogleNews & Domain-specific~\cite{won2020multimodal} \\ \hline
Chill & \begin{tabular}[c]{@{}c@{}}chilly, cold, chilled, \\chills, shivers, shiver, warm, \\frigid, frosty, balmy\end{tabular} & \begin{tabular}[c]{@{}c@{}}\textbf{chill\_out}, relax, chilled,\\ kick\_back, relaxing, \textbf{chill-out},\\ \textbf{chilled\_out}, \textbf{downtempo}, \\\textbf{down\_tempo}, unwind\end{tabular} \\ \hline
\end{tabular}
 \caption{Nearest words in GoogleNews and domain-specific word embeddings~\cite{won2020multimodal}. Music-related words are highlighted in bold.}
 \label{tab:neighbor}
\end{table}

As an alternative to using the manually annotated valence-arousal space, we can obtain tag (mood) embeddings in a more data-driven fashion. Pre-trained word embeddings, such as Word2Vec~\cite{mikolov2013efficient} and GloVe~\cite{pennington2014glove}, represent words as vectors by learning word associations from a large corpus. These embedding spaces use the cosine similarity as a measure of   semantic similarity. Recent works~\cite{choi2019zero,won2020multimodal} show the suitability of pre-trained word embedding in music retrieval and that the embedding can include more music related context by training it with music related documents~\cite{won2020multimodal,doh2020musical}---see Table~\ref{tab:neighbor}.

%%%%%%%%%%%%%%%%%%%%%%%%%%%%%%%%%%%%%%%%%%%%%%%%%%%%%%%%%%%%%%%%%%%%%%%%%%%%%%%%
%%%%%%%%%%%%%%%%%%%%%%%%%%%%%%%%%%%%%%%%%%%%%%%%%%%%%%%%%%%%%%%%%%%%%%%%%%%%%%%%
\subsection{Cross-modal Retrieval}
Instead of targeting a pre-defined embedding space, multimodal metric learning models aim at learning a shared embedding space in which semantically similar items are close together while dissimilar items are far apart in the embedding space. Unsupervised approaches leverage co-occurrence information. For example, when we collect user-created video from the web, the video and audio streams are synchronized, and this correspondence can be exploited for representation learning~\cite{arandjelovic2017look,cramer2019look}. On the other hand, supervised methods learn discriminative representations by exploiting annotated labels. Here, data from different modalities are used to train models such that data points with the same label should be close while data with different labels should be far apart. Metric learning is also used for bridging the modality gap between text and audio, such as keyword spotting~\cite{huh2021metric}, text-based audio retrieval~\cite{elizalde2019cross,oncescu2021audio}, and  tag-based music retrieval~\cite{choi2019zero,won2020multimodal} in both supervised and unsupervised ways.

Two branch metric learning~\cite{wang2018learning} is one of the most prevalent architectures for cross-modal retrieval. It consists of two branches where each branch extracts features from each modality and maps them into a shared embedding space. When optimized with a conventional triplet loss (e.g. anchor text, positive song, negative song), however, the model loses neighborhood structure within modalities. To alleviate this issue, previous work~\cite{wang2016learning} added structure-preserving constraints by using additional triplet losses within modalities (e.g., anchor text, positive text, negative text). 
% Especially, when the two modalities have different taxonomies (like our case), non-overlapped classes need to be discarded.

% Two branch metric learning~\cite{wang2018learning} is one of the most prevalent architectures for cross-modal retrieval. It consists of two branches where each branch extract features of each modality and map them into the shared embedding space. This metric learning model is optimized with triplet loss or contrastive loss that . 

\section{Models}\label{sec:model}

\begin{figure*}[ht!]
    \centering
    \includegraphics[width=1.0\linewidth]{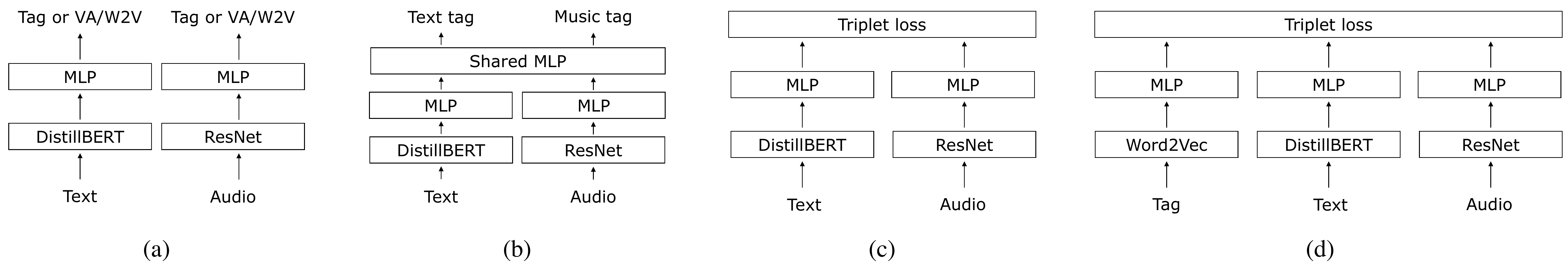} 
    \caption{Model architectures. (a) Classification and regression models (b) Multi-head classification model with shared weights (c) Two-branch metric learning (c) Three-branch metric learning.}
    \label{fig:models}
\end{figure*}

Cross-modal retrieval comprises two parts: feature extraction and bridging the modality gap.
Our text and music embeddings, $E_{text}$ and $E_{music}$ respectively, are defined as follows:
\begin{equation}
  \begin{array}{c}
E_{text} = M(P_{text}(x))\\
E_{music} = M(P_{music}(x))
  \end{array}
\end{equation}
where $P$ is a pre-trained model to extract features from each modality and $M$ is a multilayer perceptron (MLP) to map them to a multimodal embedding space. 

\subsection{Pre-trained Models for Feature Extraction}
% We use DistilBERT~\cite{sanh2019distilbert} as our text representation model $P_{text}$. It is a transformer variant that can handle multiple sentences in a compact but powerful way. 
In our work, we leverage the DistilBERT~\cite{sanh2019distilbert} transformer model for text analysis, which is a compact variant of the popular BERT transformer model~\cite{devlin2018bert, sanh2019distilbert}. We use a pre-trained model from the Huggingface library~\cite{wolf-etal-2020-transformers}.

For the music representation model $P_{music}$, we use a CNN with residual connections that are trained with mel-spectrograms  (ResNet)~\cite{won2020evaluation}. Due to its simplicity and high performance, it is a broadly used architecture not only in music but also in general audio representation learning. Our ResNet consists of 7 convolutional layers with $3\times3$ filters followed by $2\times2$ max-pooling. The model is pretrained with the MagnaTagATune dataset\footnote{We use the pre-trained model from this open source repository: https://github.com/minzwon/sota-music-tagging-models}~\cite{law2009evaluation}. 
Both pre-trained models are updated during the training process so that they can adapt to the data.

\subsection{Embedding Models to Bridge the Modality Gap}
\subsubsection{Classification}
As a starting point, we train two separate mood classification models for text and music (Figure~\ref{fig:models}-(a)). Then the model returns mood predictions and their likelihood with softmax. From the predicted text mood, songs are re-ranked based on their likelihoods of the text mood. However, this classification approach has an inherent limitation- the model cannot bridge the modalities when they have different mood taxonomies.

\subsubsection{Multi-head Classification with Shared Weights}
Multi-head model is similar to the classification model but it shares a 3-layered MLP for multimodal fusion in it (Figure~\ref{fig:models}-(b)). Since the model shares the weights across different modalities, it can predict the mood in different taxonomies by switching the classification head. We included this model to see if the shared MLP can generalize across modalities.

\subsubsection{Regression}
Following previous work~\cite{delbouys2018music}, we reformulate the classification task as a regression problem. By using NRC VAD Lexicon~\cite{mohammad2018obtaining}, emotion labels can be mapped to the valence-arousal space. However, this mapping process is hand-crafted and also they cannot handle bi-grams or tri-grams since the lexicon was created in a word-level. 
In addition to leveraging the valence-arousal space, we also experiment with a Word2Vec~\cite{mikolov2013efficient} embedding which was pre-trained with music related text~\cite{won2020multimodal}. This data-driven space supports a larger vocabulary, including bi-grams and tri-grams, and is thus more flexible.

Regression models are trained separately for each modality (Figure~\ref{fig:models}-(a)). Then the nearest items are retrieved based on their distance in the embedding space. Note that, distance metrics are Euclidean distance for the valence-arousal space, and cosine distance for the Word2Vec space. 
However, regression is a one-way optimization, i.e., optimizing text or mood into the pre-defined word embedding space. In this case, neighborhood structure within each modality can be ignored. For example, music with \textit{angry} and \textit{exciting} can share similar acoustic characteristics. However, if two words are far apart in Word2Vec space, this similarity cannot be considered by regression. This obstacle motivates us to learn a shared embedding space in a data-driven fashion using metric learning.

% Here, we use two different types of embedding spaces. One is manually annotated valence-arousal space by using NRC-VAD lexicon~\cite{mohammad2018obtaining}. And another is, more data-driven, Word2Vec~\cite{mikolov2013efficient} embedding which was pre-trained with music related text~\cite{won2020multimodal}. Valence-arousal showed it's suitability for emotion recognition, however, we cannot use theme-related tags such as \textit{party} while Word2Vec is more flexible with a larger vocabulary.

% Mood or emotion distribution is not discrete but linear. So people worked on Valence-Arousal regression. 

% But NRC-VAD relies on human annotation. W2V is more data-driven approach. So we also try W2V regression.

% \subsubsection{Regression with Shared Weights}
% From a similar motivation of the multi-head classification model, we investigate regression models with shared weights (Figure~\ref{fig:models}-(c)). But in this case, there's no multiple heads since both modalities are mapped to a shared embedding space. What we intended here is to see if the model can generalize to unseen emotions of another modality.

\subsubsection{Metric Learning}\label{subsubsec:metric}
Finally, we explore metric learning, which is a fully data-driven approach that solves the cross modal text-to-music retrieval in an end-to-end manner. Metric learning is optimized to minimize a triplet loss $\mathcal{T}$: 
\begin{equation}
  \begin{array}{c}
    \mathcal{T}(E_a, E_p, E_n) = [D(E_a, E_p) - D(E_a, E_n) + \delta]_+
  \end{array}
\end{equation}
where $D$ is a cosine distance function, $\delta$ is a margin, and $E_a$, $E_p$, $E_n$ are embedding of anchor, positive, and negative examples, respectively. $[\cdot]_+$ is rectified linear unit. Following conventional metric learning models for cross-modal retrieval, we implement a two-branch metric learning model~\cite{wang2018learning} (Figure~\ref{fig:models}-(c)) that optimizes the loss function $L$,
\begin{equation}
  \begin{array}{c}
L = \mathcal{T}(E_{text}^a, E_{music}^p, E_{music}^n).
  \end{array}
\end{equation}
% Hereinafter, we call this model as metric learning 2B where 2B stands for two branches.

However, with the triplet function, neighborhood structure or data distribution within modalities can be lost. Structure-preserving constraints~\cite{wang2016learning} can alleviate the issue but our problem is different from the case, since we have different taxonomies across the modality which includes many non-overlapped moods.

To take advantage of different mood distribution of different modalities, we investigate metric learning model with three branches (Figure~\ref{fig:models}-(d)) that results in three triplet loss functions. Each loss function is designed to optimize tag-to-text, tag-to-music, and text-to-music triplet losses as following:

\begin{equation}
  \begin{array}{c}
L_{text} = \mathcal{T}(E_{tag}^a, E_{text}^p, E_{text}^n),\\
L_{music} = \mathcal{T}(E_{tag}^a, E_{music}^p, E_{music}^n),\\
L_{cross} = \mathcal{T}(E_{text}^a, E_{music}^p, E_{music}^n).
  \end{array}
\end{equation}
The model learns a shared mood space between Word2Vec embedding and text embedding with a loss $L_{text}$, and a shared mood space between Word2Vec embedding and music embedding with a loss $L_{music}$. Finally, they are bridged together with a cross-modal triplet loss $L_{cross}$. We refer to this model as three-branch metric learning.

Since text and music have different vocabularies in our scenario, for both two-branch and three-branch metric learning, we regard the nearest tags in pre-trained Word2Vec space as positive pairs in cross-modal triplet sampling (Table~\ref{tab:similar}). We used distance-weighted sampling~\cite{wu2017sampling} for more efficient negative mining following previous work~\cite{won2020multimodal}.

\section{Experimental Design}\label{sec:exp}

\subsection{Text Datasets}
% \subsubsection{Alm's dataset}
Alm's affect dataset~\cite{alm2008affect} includes 1,383 sentences collected from books written by three different authors: B. Potter, H.C. Andersen, and the Brothers Grimm. 1,207 sentences in the dataset are annotated with one representative emotion among five: \textit{angry}, \textit{fearful}, \textit{happy}, \textit{sad}, and \textit{surprised}. To avoid unintended information leakage, we decided to split data in an author-level. 1,040 sentences by the Brothers Grimm and H.C. Andersen were used for training and 167 sentences by B. Potter were used for validation and test. 
% To estimate the performance of the pretrained model, we run a mood classification experiment and it reported 00 classification accuracy. We simply added a fully-connected layer above the pretrained model and it is updated during the training.

% \subsubsection{ISEAR dataset}
ISEAR dataset~\cite{scherer1994evidence} is a corpus with 7,666 sentences that are categorized into one of seven emotion: \textit{anger}, \textit{disgust}, \textit{fear}, \textit{joy}, \textit{sadness}, \textit{shame}, and \textit{guilt}. Each sentence describes certain antecedents and those are associated with according reactions (emotions). We split the dataset in a stratified manner with ratio of 70\% train, 15\% validation, and 15\% test set.

\subsection{Music Dataset}
% \subsubsection{Dataset selection}
There are multiple datasets for music emotion recognition such as the Million Song Dataset (MSD) subset~\cite{hu2009lyric}, the MTG-Jamendo mood subset~\cite{bogdanov2019mtg}, and the AudioSet mood subset~\cite{gemmeke2017audio}. 
% After a careful exploration of the datasets, we use the AudioSet mood subset. 
Before we choose our dataset, we run classification experiments for each subset. AudioSet subset returned the highest accuracy, which means the labeled emotions are predictable with our ResNet model. One possible reason for this result is that unlike other datasets, emotion labels of AudioSet subset are exclusive, having a single emotion label per song. This is also beneficial since we can map each song directly to the valence-arousal space or word embedding space using emotion lexicons or Word2Vec model, respectively. Otherwise, to handle multiple tags, we need to average their embedding vectors as previous researchers did~\cite{delbouys2018music}. For these simplicity and reliability reasons, we use AudioSet mood subset.

% \subsubsection{AudioSet mood subset}
AudioSet~\cite{gemmeke2017audio} mood subset consists of 16,995 music clips collected from YouTube and each audio clip is 10-second long. The dataset is categorized into 7 mood categories: \textit{happy}, \textit{funny}, \textit{sad}, \textit{tender}, \textit{exciting}, \textit{angry}, and \textit{scary}. The dataset is provided with a training set of 16,104 clips and an evaluation set of 540 clips. 

\begin{table}
\centering
\footnotesize
\begin{tabular}{@{}c|c|c|c@{}}
\toprule
Original        & VA    & W2V           & Manual \\ \midrule
anger       & angry     & angry         & angry\\
fearful     & sad       & scary         & scary\\
happy       & happy     & happy         & exciting, funny, happy\\
sad         & sad       & sad           & sad\\
surprised   & exciting  & happy         & exciting\\ \midrule
anger       & angry     & angry         & angry\\
disgust     & angry     & angry         & angry, scary\\
fear        & angry     & angry         & scary\\
guilt       & sad       & angry         & angry, sad\\
joy         & exciting  & tender        & exciting, funny, happy\\
sadness     & sad       & tender        & sad\\
shame       & angry     & sad           & angry, sad\\

\bottomrule
\end{tabular}
\caption{Similar moods from Alm's dataset (upper) and ISEAR dataset (lower). Original is from text mood taxonomy and mapped tags are from music dataset.}
\label{tab:similar}
% \end{table*}
\end{table}

\begin{table*}[ht!]
\centering
\footnotesize
\begin{tabular}{@{}l|cc|cc|cc|cc|cc|cc@{}}
\toprule
\multicolumn{1}{c}{\multirow{3}{*}{Methods}} & \multicolumn{6}{c}{Alm's dataset} &  \multicolumn{6}{c}{ISEAR dataset} \\ \cmidrule(l){2-13} 
\multicolumn{1}{c}{} & \multicolumn{2}{c}{VA}& \multicolumn{2}{c}{W2V}& \multicolumn{2}{c}{Manual}& \multicolumn{2}{c}{VA}& \multicolumn{2}{c}{W2V}& \multicolumn{2}{c}{Manual}\\ \cmidrule(l){2-13} 
\multicolumn{1}{c}{}                         & P@5 & MRR & P@5 & MRR & P@5 & MRR & P@5 & MRR & P@5 & MRR & P@5 & MRR      \\ \midrule
Classification           & 0.2161 & 0.2436  & 0.1861  & 0.2157  & 0.2161 & 0.2436  & 0.0000  & 0.0000 & 0.0000 & 0.0000  & 0.0000  & 0.0000   \\
Multi-head Classification  & 0.2819 & 0.4181  & 0.1271  & 0.1381  & 0.3446 & 0.5304  & \textbf{0.3440}  & 0.5084 & 0.3325 & 0.3625  & 0.3551  & 0.4803    \\
V-A Regression           & \textbf{0.4325} & \textbf{0.6282}  & 0.4125  & 0.5749  & \textbf{0.6100} & \textbf{0.7398}  & 0.3018  & \textbf{0.5247} & 0.1866 & 0.3709  & \textbf{0.6218}  & 0.7075    \\
% V-A Regression (shared)  & 0.3629 & 0.4978  & 0.2551  & 0.3540  & 0.4192 & 0.5234  & 0.1951  & 0.2682 & 0.1614 & 0.2274  & 0.4031  & 0.5109    \\
W2V Regression           & 0.3960 & 0.5010  & 0.4613  & 0.5591  & 0.5413 & 0.6363  & 0.3008  & 0.3829 & 0.4164 & 0.4908  & 0.5527  & \textbf{0.7668}    \\
% W2V Regression (shared)  & 0.3976 & 0.4817  & \textbf{0.5126}  & \textbf{0.6299}  & 0.5018 & 0.6487  & 0.1617  & 0.3197 & 0.2210 & 0.4771  & 0.4283  & 0.5989    \\
Metric Learning (2-branch)          & 0.3399 & 0.3778  & 0.4897  & 0.5239  & 0.5374 & 0.5579  & 0.2695  & 0.3287 & 0.3951 & 0.4336  & 0.4438  & 0.6175    \\
Metric Learning (3-branch)          & 0.3574 & 0.4348  & \textbf{0.5095}  & \textbf{0.5863}  & 0.5156 & 0.5880  & 0.2591  & 0.3445 & \textbf{0.4317} & \textbf{0.4953}  & 0.6019  & 0.6675    \\

\bottomrule
\end{tabular}
\caption{Retrieval scores}
\label{tab:mainresults}
\end{table*}

\subsection{Evaluation}
We use two evaluation metrics: Precision at 5 (P@5) and Mean Reciprocal Rank (MRR). However, since our text and audio datasets use different taxonomies, we need a mapping between the different vocabularies in order to compute the metrics directly. Thus, we map the text emotion taxonomy to the music emotion taxonomy --- see Table~\ref{tab:similar}. We introduce three possible mappings: (1) mapping based on the Euclidean distance between emotion labels in the valence-arousal space (VA), (2) the cosine distance between emotion labels in Word2Vec space (W2V), or (3) direct manual mapping of emotion labels. Given these mappings, we compute P@5 and MRR. Another challenge is the label distribution in our datasets, which is unbalanced. This can lead to over-optimistic results if the model performs well on the majority class, even if it performs very poorly on less common labels in the test dataset.
% In original to valence-arousal mapping with ISEAR dataset (lower part of Table~\ref{tab:similar}), for example, every emotions from original text are mapped to \textit{angry} or \textit{sad} except \textit{joy}. This means when a model retrieves \textit{angry} or \textit{sad} music all the time, evaluation metric will still return overoptimistic scores. 
To alleviate this problem, we compute the macro-P@5 and macro-MRR, i.e., we compute the metrics per class (emotion label) then average the per-class results. Henceforth we will use P@5 and MRR to denote \emph{macro} P@5 and MRR, respectively.

Regression models are optimized to reduce mean squared error and metric learning models are optimized with the triplet losses detailed in Section~\ref{subsubsec:metric}. We use the Adam optimizer with learning rate 0.0001 for all models. Audio inputs are resampled into 16 kHz and then converted to 128-bin mel-spectrograms via a 512-point FFT with 50\% frame overlap. Implementation details are available online~\footnote{https://github.com/minzwon/text2music-emotion-embedding.git}.

% \subsection{Optimization}

\section{Results}\label{sec:results}

\subsection{Quantitative Results}
% \subsubsection{Classification and multi-head classification}
The retrieval results for the different proposed models, using our three different proposed vocabulary mappings (VA, W2V, Manual), for our two text datasets, are presented in Table~\ref{tab:mainresults}. First, we see that the classification model fails in cross-modal retrieval. Since there are only two emotions in common between Alm's dataset and AudioSet (i.e., \textit{happy} and \textit{sad}), text inputs with other emotions will not have any retrieval result. Furthermore, there's no common emotion between ISEAR dataset and AudioSet, hence P@5 and MRR are zero in this case. Classification models can be powerful when there are exactly identical or partially overlapped vocabularies, but since it is less likely in real-world data, classification approach is less desirable for cross-modal retrieval.

% confusion matrix
% \begin{figure}
%  \centerline{
%  \includegraphics[width=0.70\columnwidth]{draft/figs/cm.pdf}}
%  \caption{Confusion matrix of a multi-head classification model trained with ISEAR dataset.}
%  \label{fig:cm}
% \end{figure}

The multi-head classification model also performs worse than other regression and metric learning models. Some metrics look optimistic but when we check the confusion matrix of the multi-head classification model, it constantly predicts one or two specific emotions (e.g., predict \textit{angry} for any type of input) no matter what the input is.  This means the shared MLP cannot generalize across different modality heads.

% Multi-head classification model cannot generalize to different head's mood distribution even if it shares the MLP across different modalities.

The regression model using valence-arousal consistently shows the best metrics as already proven in previous single-modality emotion recognition works~\cite{schmidt2010feature,han2009smers}. Since the space is carefully designed and the tag-to-space mapping process has been done manually~\cite{mohammad2018obtaining}, the valence-arousal regression suits our cross-modal retrieval task. However, this method cannot generalize to other datasets that possibly have some tags that do not have manual tag-to-space mapping. Word2Vec regression is suitable in that case. It shows slightly lower but comparable retrieval performance and it can handle abundant vocabulary, even bi-grams and tri-grams, without a manual mapping process.

% But also, a Word2Vec regression model reports comparable results. Wor2Vec embedding is a more data-driven approach that does not need manual annotation. 
% Also, it is more flexible that has capability of handling bi-grams and tri-grams.

% Different from manually annotated valence-arousal Word2Vec is a more data-driven embedding with flexibility to handle bi-grams and tri-grams. However, this regression is a one-way optimization, i.e., optimizing text or mood into the pre-defined embedding space. In this case, neighbor structure within each modality can be ignored. For example, music with \textit{angry} and \textit{exciting} can share similar acoustic characteristics. However, if two words are far apart in Word2Vec space, this similarity cannot be considered by regression.

Finally, we assess the performance of metric learning. Instead of predicting manually defined or pre-trained embeddings, metric learning aims at learning a shared embedding space across different modalities. Both two-branch and three-branch approaches claim their suitability for cross-modal retrieval, and the three-branch metric learning model consistently outperforms the two-branch model by leveraging the relationship of tag-to-text and tag-to-music within each modality.

% a learned space metric learning models with two branches and three branches show their suitability for solving cross-modal retrieval problems in an end-to-end scheme. Especially, three-branch metric learning model consistently outperforms two-branch metric learning model by leveraging relationship of tag-to-text and tag-to-music within each modality.

% reprehenderit in voluptate velit esse cillum dolore eu fugiat nulla pariatur. Excepteur sint occaecat cupidatat non proident, sunt in culpa qui officia
% commodo consequat. Duis aute irure dolor in reprehenderit in voluptate velit esse cillum dolore eu fugiat nulla pariatur. Excepteur sint occaecat cupidatat non proident, sunt in culpa qui officia
% commodo consequat. Duis aute irure dolor in reprehenderit in voluptate velit esse cillum dolore eu fugiat nulla pariatur. Excepteur sint occaecat cupidatat non proident, sunt in culpa qui officia
% commodo consequat. Duis aute irure dolor in reprehenderit in voluptate velit esse cillum dolore eu fugiat nulla pariatur. Excepteur sint occaecat cupidatat non proident, sunt in culpa qui officia
% commodo consequat. Duis aute irure dolor in reprehenderit in voluptate velit esse cillum dolore eu fugiat nulla pariatur. Excepteur sint occaecat cupidatat non proident, sunt in culpa qui officia

% qualitative
\subsection{Qualitative Results}
\begin{figure}[t]
    \centering
    \includegraphics[width=1.0\linewidth]{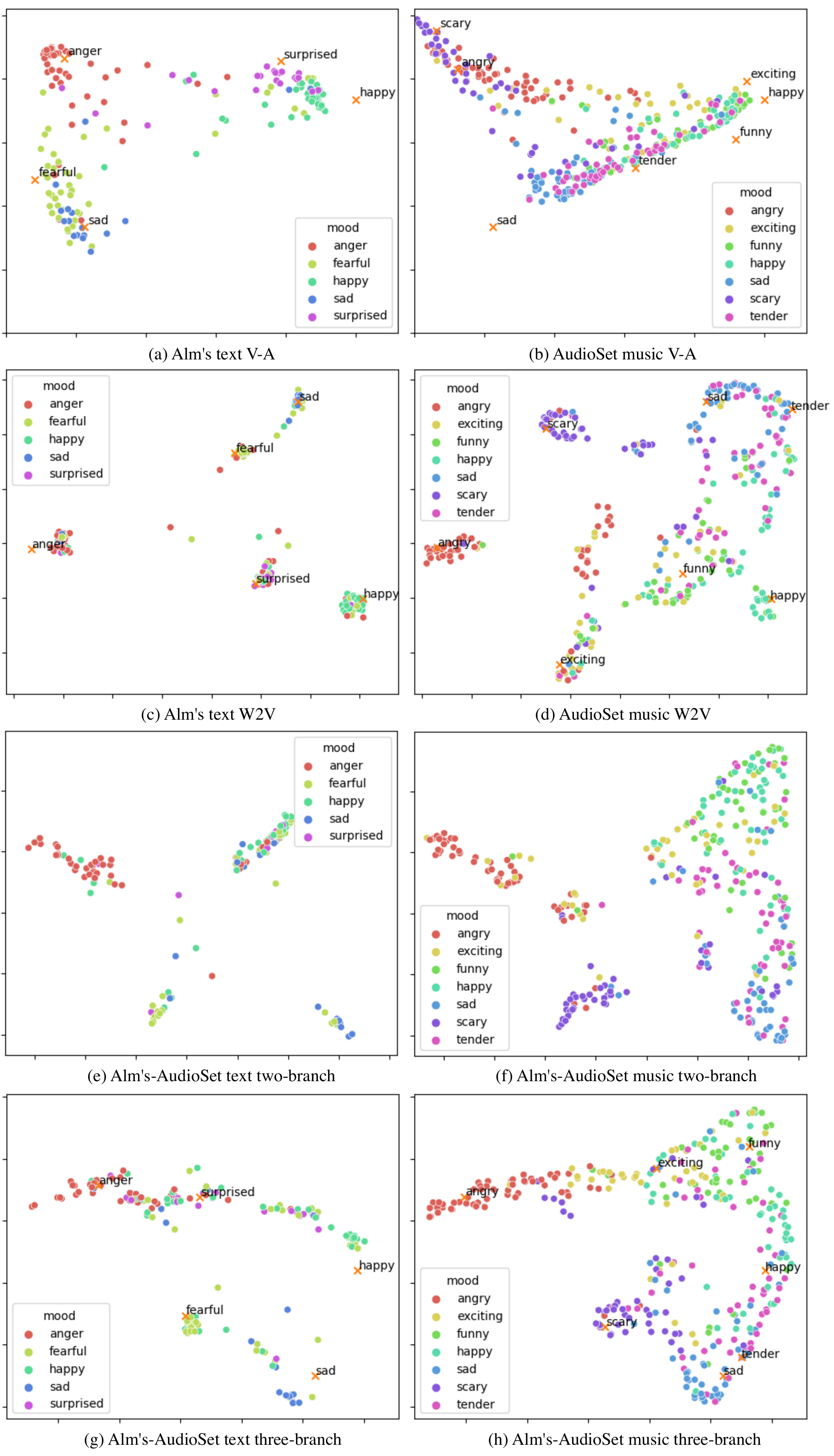} 
    \caption{Valence-arousal embedding (first row), UMAP of Word2Vec embedding (second row), UMAP of shared embedding space from two-branch metric learning (third row), and UMAP of shared embedding space from three-branch metric learning (fourth row).  }
    \label{fig:embedding}
\end{figure}

To further investigate the characteristics of various embedding spaces, we visualize them with 2D projection---Figure~\ref{fig:embedding}. Due to limited space, we only visualize embedding spaces with Alm's dataset and AudioSet mood subset. Note that they are all predicted embeddings using the test set. Except valence-arousal space (first row), which is already 2D, high dimensional embedding spaces are projected to a 2D space using the uniform manifold approximation and projection (UMAP)~\cite{mcinnes2018umap}. We use UMAP since it preserves more of the global structure compared to tSNE~\cite{van2008visualizing}. In the projection process, we first fit the UMAP with one modality (in our figure: music), then projected other embeddings (in our figure: tag and text) into the fitted 2D space.

First of all, for both the Word2Vec embedding space and the metric learning space, relevant moods from different taxonomies are neighboring together in the embedding space. This is natural for the Word2Vec space because each modality is fitted to optimize the pre-defined word embeddings. But this neighboring also can be found in metric learning space. In Figure~\ref{fig:embedding}-(g) and (h) for example, \textit{anger} from text and \textit{angry} from music are together, and \textit{fearful} from text and \textit{scary} from music are together. Note that Figure~\ref{fig:embedding}-(e) and (f) do not have word embeddings since the two-branch metric learning model does not have a branch to map the mood tags into the embedding space.

One of our main motivations to use metric learning with three branches is to preserve neighborhood structure within modalities. Since Word2Vec regression is a one-way optimization, their embeddings are very discriminative (Figure~\ref{fig:embedding}-(c)). Also, the two-branch neural network does not have any means to learn the neighborhood structure of each modality. Especially, as shown in Table~\ref{tab:similar}, when two-branch metric learning uses the mapping of Alm's mood into AudioSet mood with Word2Vec similarity, \textit{exciting} and \textit{tender} from music are not being used in training. If we compare Figure~\ref{fig:embedding}-(f) and (h), \textit{exciting} music in (h) are more continuously distributed between \textit{angry} and \textit{happy} while they are simply with \textit{happy} in (f). Also, when we compare text embeddings (see (e) and (g)), \textit{surprised} is continuously distributed between \textit{anger} and \textit{happy} in (g) but not in (e). This continuity between music and text can be found in the manually annotated valence-arousal space (see (b) and (a), respectively), which means the proposed three-branch metric learning model preserves neighborhood structure within modalities in the learned multi-modal embedding space. We summarize all the introduced characteristics in Table~\ref{tab:summary}.

\begin{table}
\centering
\footnotesize

\begin{tabular}{@{}l|c|c|c@{}}
% \raggedleft
\toprule
Model        & Retrieval    & Distribution  & Mapping\\ \midrule
Classification                  & fail       & .         & .\\
Multi-head classification       & fail       & .         & .\\
V-A regression                  & success    & continuous  & manual\\
W2V regression                  & success    & discriminative  & data-driven\\
Metric learning (2 branch)              & success    & discriminative  & data-driven\\
Metric learning (3 branch)              & success    & continuous  & data-driven\\
\bottomrule
\end{tabular}
\caption{Characteristics of different models}
\label{tab:summary}
% \end{table*}
\end{table}

\section{Conclusion}\label{sec:conclusion}
In this work we tackled the task of matching music to text with the goal of allowing users to enhance their text-based stories with music that matches the mood of the text. 
% formulation and challenge
We formulated the problem as a cross-modal text-to-music retrieval problem, and identified the lack of a shared vocabulary as a key challenge for bridging the gap between modalities.
% what we propose
To address this challenge, we proposed and investigated several emotion embedding spaces, both manually defined (valence/arousal) and data-driven (Word2Vec and metric learning), to bridge between the text and music modalities. 
% results
Our experiments showed that by leveraging these embedding spaces, we were able to facilitate cross modal retrieval successfully. We showed that the carefully designed valence-arousal space can bridge different modalities, but this can be also achieved via data-driven embedding spaces. Especially, our proposed three-branch metric learning model preserves the neighborhood structure of emotions within modalities.
% We showed that our method can leverage the well established valence-arousal space to bridge between modalities, but that it can also achieve this goal via data-driven embedding spaces. 
By leveraging data-driven embeddings, our approach has the potential of being generalized to other cross-modal retrieval tasks that require broader or completely different vocabularies.

\section{Acknowledgement}\label{sec:acknowledgement}
This work was funded by the predoctoral grant MDM-2015-0502-17-2 from the Spanish Ministry of Economy and Competitiveness linked to the Maria de Maeztu Units of Excellence Programme (MDM-2015-0502).

\bibliography{main}

\begin{thebibliography}{10}
\providecommand{\url}[1]{#1}
\csname url@samestyle\endcsname
\providecommand{\newblock}{\relax}
\providecommand{\bibinfo}[2]{#2}
\providecommand{\BIBentrySTDinterwordspacing}{\spaceskip=0pt\relax}
\providecommand{\BIBentryALTinterwordstretchfactor}{4}
\providecommand{\BIBentryALTinterwordspacing}{\spaceskip=\fontdimen2\font plus
\BIBentryALTinterwordstretchfactor\fontdimen3\font minus
  \fontdimen4\font\relax}
\providecommand{\BIBforeignlanguage}[2]{{%
\expandafter\ifx\csname l@#1\endcsname\relax
\typeout{** WARNING: IEEEtran.bst: No hyphenation pattern has been}%
\typeout{** loaded for the language `#1'. Using the pattern for}%
\typeout{** the default language instead.}%
\else
\language=\csname l@#1\endcsname
\fi
#2}}
\providecommand{\BIBdecl}{\relax}
\BIBdecl

\bibitem{kim2010music}
Y.~E. Kim, E.~M. Schmidt, R.~Migneco, B.~G. Morton, P.~Richardson, J.~Scott,
  J.~A. Speck, and D.~Turnbull, ``Music emotion recognition: A state of the art
  review,'' in \emph{Proc. of International Society for Music Information
  Retrieval Conference (ISMIR)}, 2010.

\bibitem{wang2017adversarial}
B.~Wang, Y.~Yang, X.~Xu, A.~Hanjalic, and H.~T. Shen, ``Adversarial cross-modal
  retrieval,'' in \emph{Proc. of the 25th ACM International Conference on
  Multimedia}, 2017.

\bibitem{yao2015learning}
T.~Yao, T.~Mei, and C.-W. Ngo, ``Learning query and image similarities with
  ranking canonical correlation analysis,'' in \emph{Proc. of the IEEE
  International Conference on Computer Vision (ICCV)}, 2015.

\bibitem{zhen2019deep}
L.~Zhen, P.~Hu, X.~Wang, and D.~Peng, ``Deep supervised cross-modal
  retrieval,'' in \emph{Proc. of the IEEE/CVF Conference on Computer Vision and
  Pattern Recognition (CVPR)}, 2019.

\bibitem{wei2016cross}
Y.~Wei, Y.~Zhao, C.~Lu, S.~Wei, L.~Liu, Z.~Zhu, and S.~Yan, ``Cross-modal
  retrieval with cnn visual features: A new baseline,'' \emph{IEEE Transactions
  on Cybernetics}, vol.~47, no.~2, 2016.

\bibitem{wang2016comprehensive}
K.~Wang, Q.~Yin, W.~Wang, S.~Wu, and L.~Wang, ``A comprehensive survey on
  cross-modal retrieval,'' \emph{arXiv preprint arXiv:1607.06215}, 2016.

\bibitem{mohammad2013crowdsourcing}
S.~M. Mohammad and P.~D. Turney, ``Crowdsourcing a word--emotion association
  lexicon,'' \emph{Computational Intelligence}, vol.~29, no.~3, 2013.

\bibitem{strapparava2004wordnet}
C.~Strapparava, A.~Valitutti \emph{et~al.}, ``Wordnet affect: an affective
  extension of wordnet.'' in \emph{Proc. of International Conference on
  Language Resources and Evaluation}, 2004.

\bibitem{danisman2008feeler}
T.~Danisman and A.~Alpkocak, ``Feeler: Emotion classification of text using
  vector space model,'' in \emph{AISB Convention: Communication, Interaction
  and Social Intelligence}, vol.~1, 2008.

\bibitem{hasan2014emotex}
M.~Hasan, E.~Rundensteiner, and E.~Agu, ``Emotex: Detecting emotions in twitter
  messages,'' 2014.

\bibitem{abdul2017emonet}
M.~Abdul-Mageed and L.~Ungar, ``Emonet: Fine-grained emotion detection with
  gated recurrent neural networks,'' in \emph{Proc. of annual meeting of the
  association for computational linguistics}, 2017.

\bibitem{batbaatar2019semantic}
E.~Batbaatar, M.~Li, and K.~H. Ryu, ``Semantic-emotion neural network for
  emotion recognition from text,'' \emph{IEEE Access}, vol.~7, 2019.

\bibitem{cortiz2021exploring}
D.~Cortiz, ``Exploring transformers in emotion recognition: a comparison of
  bert, distillbert, roberta, xlnet and electra,'' \emph{arXiv preprint
  arXiv:2104.02041}, 2021.

\bibitem{devlin2018bert}
J.~Devlin, M.-W. Chang, K.~Lee, and K.~Toutanova, ``Bert: Pre-training of deep
  bidirectional transformers for language understanding,'' \emph{Proc. of
  Conference of the North {A}merican Chapter of the Association for
  Computational Linguistics: Human Language Technologies, Volume 1}, 2019.

\bibitem{liu2019roberta}
Y.~Liu, M.~Ott, N.~Goyal, J.~Du, M.~Joshi, D.~Chen, O.~Levy, M.~Lewis,
  L.~Zettlemoyer, and V.~Stoyanov, ``Roberta: A robustly optimized bert
  pretraining approach,'' \emph{arXiv preprint arXiv:1907.11692}, 2019.

\bibitem{sanh2019distilbert}
V.~Sanh, L.~Debut, J.~Chaumond, and T.~Wolf, ``Distilbert, a distilled version
  of bert: smaller, faster, cheaper and lighter,'' \emph{Neural Information
  Processing Systems Workshop on Energy Efficient Machine Learning and
  Cognitive Computing}, 2019.

\bibitem{tzanetakis2007marsyas}
G.~Tzanetakis, ``Marsyas submissions to mirex 2007,'' \emph{Music Information
  Retrieval Evaluation eXchange}, 2007.

\bibitem{peeters2008generic}
G.~Peeters, ``A generic training and classification system for mirex08
  classification tasks: audio music mood, audio genre, audio artist and audio
  tag,'' in \emph{Proc. of the International Symposium on Music Information
  Retrieval (ISMIR)}, 2008.

\bibitem{cao2009thinkit}
C.~Cao and M.~Li, ``Thinkit’s submissions for mirex2009 audio music
  classification and similarity tasks,'' \emph{Music Information Retrieval
  Evaluation eXchange}, 2009.

\bibitem{lidy2016parallel}
T.~Lidy, A.~Schindler \emph{et~al.}, ``Parallel convolutional neural networks
  for music genre and mood classification,'' \emph{Music Information Retrieval
  Evaluation eXchange}, 2016.

\bibitem{delbouys2018music}
R.~Delbouys, R.~Hennequin, F.~Piccoli, J.~Royo-Letelier, and M.~Moussallam,
  ``Music mood detection based on audio and lyrics with deep neural net,''
  \emph{In Proc. of International Society for Music Information Retrieval
  Conference (ISMIR)}, 2018.

\bibitem{choi2016automatic}
K.~Choi, G.~Fazekas, and M.~Sandler, ``Automatic tagging using deep
  convolutional neural networks,'' \emph{In Proc. of International Society for
  Music Information Retrieval Conference (ISMIR)}, 2016.

\bibitem{lee2017sample}
J.~Lee, J.~Park, K.~L. Kim, and J.~Nam, ``Sample-level deep convolutional
  neural networks for music auto-tagging using raw waveforms,'' \emph{In Proc.
  of Sound and music computing (SMC)}, 2017.

\bibitem{pons2018end}
J.~Pons, O.~Nieto, M.~Prockup, E.~Schmidt, A.~Ehmann, and X.~Serra,
  ``End-to-end learning for music audio tagging at scale,'' \emph{In Proc. of
  International Society for Music Information Retrieval Conference (ISMIR)},
  2018.

\bibitem{won2020data}
M.~Won, S.~Chun, , O.~Nieto, and X.~Serra, ``Data-driven harmonic filters for
  audio representation learning,'' \emph{In Proc. of International Conference
  on Acoustics, Speech and Signal Processing (ICASSP)}, 2020.

\bibitem{lee2020metric}
J.~Lee, N.~J. Bryan, J.~Salamon, Z.~Jin, and J.~Nam, ``Metric learning vs
  classification for disentangled music representation learning,'' in \emph{In
  Proc. of International Society for Music Information Retrieval Conference
  (ISMIR)}, 2020.

\bibitem{won2020evaluation}
M.~Won, A.~Ferraro, D.~Bogdanov, and X.~Serra, ``Evaluation of cnn-based
  automatic music tagging models,'' \emph{In Proc. of Sound and Music Computing
  (SMC)}, 2020.

\bibitem{lin2017focal}
T.-Y. Lin, P.~Goyal, R.~Girshick, K.~He, and P.~Doll{\'a}r, ``Focal loss for
  dense object detection,'' in \emph{Proc. of the IEEE International Conference
  on Computer Vision (ICCV)}, 2017.

\bibitem{knox2020media}
``Mediaeval 2020 emotion and theme recognition in music task: Loss function
  approaches for multi-label music tagging,'' \emph{MediaEval2020}, 2020.

\bibitem{schmidt2010feature}
E.~M. Schmidt, D.~Turnbull, and Y.~E. Kim, ``Feature selection for
  content-based, time-varying musical emotion regression,'' in \emph{Proc. of
  the international conference on Multimedia information retrieval}, 2010, pp.
  267--274.

\bibitem{han2009smers}
B.-j. Han, S.~Rho, R.~B. Dannenberg, and E.~Hwang, ``Smers: Music emotion
  recognition using support vector regression.'' in \emph{In Proc. of
  International Society for Music Information Retrieval Conference (ISMIR)},
  2009.

\bibitem{russell1980circumplex}
J.~A. Russell, ``A circumplex model of affect.'' \emph{Journal of personality
  and social psychology}, vol.~39, no.~6, p. 1161, 1980.

\bibitem{thayer1990biopsychology}
R.~E. Thayer, \emph{The biopsychology of mood and arousal}.\hskip 1em plus
  0.5em minus 0.4em\relax Oxford University Press, 1990.

\bibitem{mohammad2018obtaining}
S.~Mohammad, ``Obtaining reliable human ratings of valence, arousal, and
  dominance for 20,000 english words,'' in \emph{Proc. of the 56th Annual
  Meeting of the Association for Computational Linguistics (Volume 1: Long
  Papers)}, 2018, pp. 174--184.

\bibitem{won2020multimodal}
M.~Won, S.~Oramas, O.~Nieto, F.~Gouyon, and X.~Serra, ``Multimodal metric
  learning for tag-based music retrieval,'' \emph{In Proc. of International
  Conference on Acoustics, Speech and Signal Processing (ICASSP)}, 2021.

\bibitem{mikolov2013efficient}
T.~Mikolov, K.~Chen, G.~Corrado, and J.~Dean, ``Efficient estimation of word
  representations in vector space,'' \emph{arXiv preprint arXiv:1301.3781},
  2013.

\bibitem{pennington2014glove}
J.~Pennington, R.~Socher, and C.~D. Manning, ``Glove: Global vectors for word
  representation,'' in \emph{Proc. of the 2014 conference on empirical methods
  in natural language processing (EMNLP)}, 2014, pp. 1532--1543.

\bibitem{choi2019zero}
J.~Choi, J.~Lee, J.~Park, and J.~Nam, ``Zero-shot learning for audio-based
  music classification and tagging,'' \emph{In Proc. of International Society
  for Music Information Retrieval Conference (ISMIR)}, 2019.

\bibitem{doh2020musical}
S.~Doh, J.~Lee, T.~H. Park, and J.~Nam, ``Musical word embedding: Bridging the
  gap between listening contexts and music,'' \emph{arXiv preprint
  arXiv:2008.01190}, 2020.

\bibitem{arandjelovic2017look}
R.~Arandjelovic and A.~Zisserman, ``Look, listen and learn,'' in \emph{Proc. of
  the IEEE International Conference on Computer Vision (ICCV)}, 2017, pp.
  609--617.

\bibitem{cramer2019look}
J.~Cramer, H.-H. Wu, J.~Salamon, and J.~P. Bello, ``Look, listen, and learn
  more: Design choices for deep audio embeddings,'' in \emph{IEEE International
  Conference on Acoustics, Speech and Signal Processing (ICASSP)}, 2019.

\bibitem{huh2021metric}
J.~Huh, M.~Lee, H.~Heo, S.~Mun, and J.~S. Chung, ``Metric learning for keyword
  spotting,'' in \emph{2021 IEEE Spoken Language Technology Workshop
  (SLT)}.\hskip 1em plus 0.5em minus 0.4em\relax IEEE, 2021, pp. 133--140.

\bibitem{elizalde2019cross}
B.~Elizalde, S.~Zarar, and B.~Raj, ``Cross modal audio search and retrieval
  with joint embeddings based on text and audio,'' in \emph{IEEE International
  Conference on Acoustics, Speech and Signal Processing (ICASSP)}, 2019.

\bibitem{oncescu2021audio}
A.-M. Oncescu, A.~Koepke, J.~F. Henriques, Z.~Akata, and S.~Albanie, ``Audio
  retrieval with natural language queries,'' \emph{arXiv e-prints}, pp.
  arXiv--2105, 2021.

\bibitem{wang2018learning}
L.~Wang, Y.~Li, J.~Huang, and S.~Lazebnik, ``Learning two-branch neural
  networks for image-text matching tasks,'' \emph{IEEE Transactions on Pattern
  Analysis and Machine Intelligence}, vol.~41, no.~2, pp. 394--407, 2018.

\bibitem{wang2016learning}
L.~Wang, Y.~Li, and S.~Lazebnik, ``Learning deep structure-preserving
  image-text embeddings,'' in \emph{Proc. of the IEEE conference on computer
  vision and pattern recognition (CVPR)}, 2016.

\bibitem{wolf-etal-2020-transformers}
T.~Wolf, L.~Debut, V.~Sanh, J.~Chaumond, C.~Delangue, A.~Moi, P.~Cistac,
  T.~Rault, R.~Louf, M.~Funtowicz, J.~Davison, S.~Shleifer, P.~von Platen,
  C.~Ma, Y.~Jernite, J.~Plu, C.~Xu, T.~L. Scao, S.~Gugger, M.~Drame, Q.~Lhoest,
  and A.~M. Rush, ``Transformers: State-of-the-art natural language
  processing,'' in \emph{Proc. of the 2020 Conference on Empirical Methods in
  Natural Language Processing: System Demonstrations}, 2020.

\bibitem{law2009evaluation}
E.~Law, K.~West, M.~I. Mandel, M.~Bay, and J.~S. Downie, ``Evaluation of
  algorithms using games: The case of music tagging.'' in \emph{In Proc. of
  International Society for Music Information Retrieval Conference (ISMIR)},
  2009.

\bibitem{wu2017sampling}
C.-Y. Wu, R.~Manmatha, A.~J. Smola, and P.~Krahenbuhl, ``Sampling matters in
  deep embedding learning,'' in \emph{Proc. of the IEEE International
  Conference on Computer Vision (ICCV)}, 2017.

\bibitem{alm2008affect}
E.~C.~O. Alm, \emph{Affect in* text and speech}.\hskip 1em plus 0.5em minus
  0.4em\relax Citeseer, 2008.

\bibitem{scherer1994evidence}
K.~R. Scherer and H.~G. Wallbott, ``Evidence for universality and cultural
  variation of differential emotion response patterning.'' \emph{Journal of
  personality and social psychology}, vol.~66, no.~2, p. 310, 1994.

\bibitem{hu2009lyric}
X.~Hu, J.~S. Downie, and A.~F. Ehmann, ``Lyric text mining in music mood
  classification,'' \emph{American music}, vol. 183, no. 5,049, pp. 2--209,
  2009.

\bibitem{bogdanov2019mtg}
D.~Bogdanov, M.~Won, P.~Tovstogan, A.~Porter, and X.~Serra, ``The mtg-jamendo
  dataset for automatic music tagging,'' \emph{Machine Learning for Music
  Discovery Workshop, International Conference on Machine Learning}, 2019.

\bibitem{gemmeke2017audio}
J.~F. Gemmeke, D.~P. Ellis, D.~Freedman, A.~Jansen, W.~Lawrence, R.~C. Moore,
  M.~Plakal, and M.~Ritter, ``Audio set: An ontology and human-labeled dataset
  for audio events,'' in \emph{IEEE International Conference on Acoustics,
  Speech and Signal Processing (ICASSP)}, 2017.

\bibitem{mcinnes2018umap}
L.~McInnes, J.~Healy, and J.~Melville, ``Umap: Uniform manifold approximation
  and projection for dimension reduction,'' \emph{arXiv preprint
  arXiv:1802.03426}, 2018.

\bibitem{van2008visualizing}
L.~Van~der Maaten and G.~Hinton, ``Visualizing data using t-sne.''
  \emph{Journal of machine learning research}, vol.~9, no.~11, 2008.

\end{thebibliography}

% For non bibtex users:
%\begin{thebibliography}{citations}
% \bibitem{Author:17}
% E.~Author and B.~Authour, ``The title of the conference paper,'' in {\em Proc.
% of the Int. Society for Music Information Retrieval Conf.}, (Suzhou, China),
% pp.~111--117, 2017.
%
% \bibitem{Someone:10}
% A.~Someone, B.~Someone, and C.~Someone, ``The title of the journal paper,''
%  {\em Journal of New Music Research}, vol.~A, pp.~111--222, September 2010.
%
% \bibitem{Person:20}
% O.~Person, {\em Title of the Book}.
% \newblock Montr\'{e}al, Canada: McGill-Queen's University Press, 2021.
%
% \bibitem{Person:09}
% F.~Person and S.~Person, ``Title of a chapter this book,'' in {\em A Book
% Containing Delightful Chapters} (A.~G. Editor, ed.), pp.~58--102, Tokyo,
% Japan: The Publisher, 2009.
%
%
%\end{thebibliography}

\end{document}